\documentclass[twocolumn]{aastex631}

\newcommand{\angstrom}{\mbox{\normalfont\AA}}
\usepackage{amsmath}
\usepackage{graphicx}
\usepackage{environ}
\NewEnviron{myequation}{%
\begin{equation}
\scalebox{0.8}{$\BODY$}
\end{equation}
}

\shorttitle{The Mg-K anti-correlation in $\omega$ Centauri}
\shortauthors{Alvarez et al.}

\graphicspath{{./}{figures/}}

\begin{document}

\title{The Mg-K anti-correlation in $\omega$ Centauri
\footnote{Based on observations collected at the European Organisation for Astronomical Research in the Southern Hemisphere under ESO programme 095.D-0539.}}

\correspondingauthor{Deimer Antonio Alvarez Garay}
\email{deimer.alvarezgaray2@unibo.it}

\author{Deimer Antonio Alvarez Garay}
\author{Alessio Mucciarelli}
\affiliation{Dipartimento di Fisica e Astronomia, Università degli Studi di Bologna, Via Gobetti 93/2, I-40129 Bologna, Italy}
\affiliation{INAF, Osservatorio di Astrofisica e Scienza dello Spazio di Bologna, Via Gobetti 93/3, I-40129 Bologna, Italy}
\author{Carmela Lardo}
\affiliation{Dipartimento di Fisica e Astronomia, Università degli Studi di Bologna, Via Gobetti 93/2, I-40129 Bologna, Italy}
\author{Michele Bellazzini}
\affiliation{INAF, Osservatorio di Astrofisica e Scienza dello Spazio di Bologna, Via Gobetti 93/3, I-40129 Bologna, Italy}
\author{Thibault Merle}
\affiliation{Institut d’Astronomie et d’Astrophysique, Université Libre de Bruxelles, CP. 226, Boulevard du Triomphe, 1050 Brussels, Belgium}

\begin{abstract}
We present [K/Fe] abundance ratios for a sample of 450 stars in $\omega$ Centauri, using high resolution spectra acquired with the multi-object spectrograph FLAMES@VLT. Abundances for Fe, Na and Mg were also derived. We detected intrinsic K variations in the analysed stars. Moreover, [K/Fe] shows a significant correlation with [Na/Fe] and anti-correlation with [Mg/Fe]. The presence of a clear-cut Mg-K anti-correlation makes $\omega$ Centauri the third stellar system, after NGC 2419 and NGC 2808, hosting a sub-population of stars with [Mg/Fe]$<0.0$ dex, K-enriched in the case of $\omega$ Centauri by $\sim0.3$ dex with respect to the Mg-rich stars ([Mg/Fe]$>0.0$ dex). The correlation/anti-correlation between K and other light elements involved in chemical anomalies supports the idea that the spread in [K/Fe] can be associated to the same self-enrichment process typical of globular clusters. We suggest that significant variations in K abundances perhaps can be found in the most massive and/or metal-poor globular clusters as manifestation of an extreme self-enrichment process. Theoretical models face problems to explain the K production in globular clusters. Indeed, models where asymptotic giant branch stars are responsible for the Mg-K anti-correlation only qualitatively agree with the observations. Finally, we discovered a peculiar star with an extraordinary K overabundance ([K/Fe]=+1.60 dex) with respect to the other stars with similar [Mg/Fe]. We suggest that this K-rich star could be formed from the pure ejecta of AGB stars before dilution with pristine material.
\end{abstract}


\section{Introduction} \label{sec:intro}
   Over the past thirty years, increasingly accurate photometric and spectroscopic studies have changed our comprehension of the stellar populations in globular clusters (GCs). The traditional picture where all the stars in a given GC have the same age and chemical composition has been replaced by a new view where multiple populations (MPs) with different light element abundances (C, N, O, Na, Mg, Al) are present in the same cluster (\citealt{carretta09}, \citealt{meszaros15}, \citealt{pancino17}). In contrast, GCs can still be considered as chemically homogeneous regarding heavy elements like iron and iron-peak elements, even if some GCs show variations in the s-elements abundances (e.g. \citealt{marino_15}).\\
   These chemical differences appear as correlations and anti-correlations between light elements and suggest a crucial role of the hot CNO cycle and its secondary chains NeNa and MgAl (\citealt{langer93}, \citealt{pran07}). The most popular theoretical models for the formation of MPs require the occurrence of subsequent episodes of star formation where second generation stars (SG) were formed out from the material polluted by stars from the first generation (FG or polluters) during the first 100-200 Myr of the cluster life. \\
   Several stellar polluters where proposed in literature- intermediate-mass asymptotic giant branch (AGB) stars (\citealt{dercole_10}), supermassive stars (\citealt{denissenkov_14}), fast rotating massive stars (FRMS; \citealt{krause_13}) and interacting binary stars (\citealt{mink_09})- essentially because they are able to reach the temperatures required to activate the CNO cycle. However, all these models have difficulties to account for all the observational evidence gathered so far (e.g. \citealt{bastian18}, \citealt{gratton_19}). Consequently, new observational clues on the nature of the polluters are highly desirable. Potassium (K) is a new entry among elements whose abundance varies within GC stars. \citet{mucc12} and \citet{cohen12} firstly discovered the presence of an extended Mg-K anti-correlation in the massive GC NGC 2419. Stars in this GC cover an unusually large range of K abundances, from solar values up to [K/Fe]$\sim +2$ dex. Magnesium also exhibits large variations, from the typical $\alpha$-enhanced values observed in other GCs down to [Mg/Fe]$\sim -1$ dex. Such magnesium depletion has not been observed up to that time neither in GCs or in the field. \\
   The Mg-K anti-correlation was detected also in NGC 2808 (\citealt{mucc15}), although less extended than that of NGC 2419. \citet{mucc15} measured the Mg and K abundances for twelve stars of NGC 2808 and among these, three Mg-poor ([Mg/Fe]$<0.0$ dex) stars show enhanced K abundances with respect to Mg-rich stars ([Mg/Fe]$>0.0$ dex).\\
   In NGC 4833, \citet{carretta21} found evidence for intrinsic K variations in a sample of 59 stars. They also found a Mg-K anti-correlation in 38 stars, even if none of them have [Mg/Fe]$<0.0$ dex.\\
   Measured K abundances show significant correlations and anti-correlations with other light elements such as O, Na and Al, hence scatter observed in [K/Fe] seems to be related to the same mechanism producing the variations in the other light elements. In particular, the Mg-K anti-correlation was detected only in very massive and/or metal-poor GCs. These would support the notion that it is a manifestation of an extreme self-enrichment process where all the CNO secondary chains are very efficient (\citealt{mucc17}). \\
   The origin of the Mg-K anti-correlation is not clear. \citet{ventura12} proposed a model where the Mg-poor/K-rich stars in NGC 2419 formed from the ejecta of AGB and super-AGB stars. In their model the consumption of Mg is accompanied not only by the production of Al, but also of K from proton-capture reaction on Ar nuclei. However, in order to reproduce the abundances observed in NGC 2419 this model requires a significant increase in the reaction cross section or in the burning temperature at the base of the envelope during the Hot Bottom Burning (HBB) phase. On the other hand, no K production is expected in FRMS or supermassive stars which are not able to reach the temperatures needed to synthesize K (\citealt{pran17}).\\\\  
   $\omega$ Centauri (NGC 5139) represents an interesting target to search for spread in [K/Fe]. It is one of the most complex stellar systems overall and the most massive among GCs with M $=(4.05\pm0.10)\cdot 10^6\;\mathrm{M_{\odot}}$ (\citealt{souza_13}). It spans a large metallicity range from [Fe/H]$\approx -2.2$ dex up to $-0.5$ dex with different peaks in the metallicity distribution (e.g. \citealt{pancino_00}, \citealt{johnson}) and it is one of the few systems hosting Mg-poor stars (\citealt{NDC95}, \citealt{meszaros20}). The observational properties of $\omega$ Centauri would suggest that it represents the remnant of an ancient nucleated dwarf galaxy that merged with Galaxy in early epochs (\citealt{bekki03}).\\
   The large metallicity spread can be attributed to multiple star formation episodes that last few Gyr (\citealt{smith_00}, \citealt{romano_10}).  However, the abundances of the light elements are more similar to those observed in the typical GCs than those of the dwarf galaxies, and include the presence of O-Na, Mg-Al, O-Al, Mg-Si anti-correlations together with Na-Al and Al-Si correlations (\citealt{NDC95}, \citealt{smith_00}, \citealt{johnson}, \citealt{marino11}, \citealt{pancino17}, \citealt{meszaros20}). The large metallicity distribution and the variations in light elements suggest that both core-collapse supernovae and the products of the proton-capture reactions played an important role in the $\omega$ Centauri chemical enrichment history. \\
   Because of its complex history of formation and evolution, $\omega$ Centauri represents an ideal candidate to search for the Mg-K anti-correlation. A hint of the presence of a Mg-K anti-correlation was found by \citet{meszaros20}. They identified seven Mg-poor stars that are slightly enriched in K compared to the Mg-rich stars in their sample. Nevertheless, the infrared lines used to derive K abundance are weak and blended, so \cite{meszaros20} concluded that a K-enhancement in the Mg-poor stars of $\omega$ Centauri cannot be convincingly asserted. \\
   In this study we investigate the presence of a Mg-K anti-correlation in $\omega$ Centauri. In Section $\ref{sec:obs}$ we present observational data and describe the analysis in Section $\ref{sec:chem}$. In Section $\ref{sec:results}$ we review the results of chemical analysis. Finally, we present our conclusions in Section $\ref{sec:disc}$. 


\section{OBSERVATIONS} \label{sec:obs}
   Observations were performed with the multi-object spectrograph FLAMES (\citealt{pasq02}), within the ESO programme 095.D-0539 (P.I. Mucciarelli). We used FLAMES in the GIRAFFE mode that allows us to allocate simultaneously up to $132$ fibers. All the targets were observed with both HR11 and HR18 setups, covering the wavelength range from $5597$ to $5840$ $\angstrom$ and from $7648$ to $7889$ $\angstrom$, and with a spectral resolution of $29500$ and $20150$, respectively. The first setup allows to measure the Mg line at $5711$ $\angstrom$ and the Na doublet at $5682$ and $5688$ $\angstrom$, the second the K I resonance line at $7699$ $\angstrom$. We checked for each target that K line was not contaminated by telluric lines. This is due to the high radial velocity of $\omega$ Centauri ($232.7\pm0.2$, $\sigma = 17.6$ km s$^{-1}$), \citealt{baumgardt_18}).
   Also, several Fe lines are included in the considered wavelength range. \\
   We selected targets among the member stars of $\omega$ Centauri already analysed in previous works (\citealt{NDC95}, \citealt{johnson}, \citealt{marino11}). Also we considered only stars that result to be not contaminated by neighbour stars within the size of the GIRAFFE fibers.  We observed a total of $450$ stars: $350$ are in common with \citet{johnson}, $85$ with \citet{marino11} and $15$ with \citet{NDC95}.\\
   Since the faintest targets have $V\sim14$, two exposures of 1300 s and two of 300 s each were sufficient to reach $SNR\sim70$ and $SNR\sim100$ for HR11 and HR18, respectively. The splitting of the observations allows to get rid of the effect of cosmic rays and other spurious transient effects. During each exposure some fibers ($\sim$15) were dedicated to observe empty sky regions to sample the sky background.\\
   Spectra were reduced with the GIRAFFE ESO pipeline\footnote{\url{https://www.eso.org/sci/software/pipelines/giraffe/giraffe-pipe-recipes.html}}, that includes bias subtraction, flat field correction, wavelength calibration and spectral extraction. For each exposure, the individual sky spectra were median-combined together and the resulting spectrum was subtracted from each stellar spectrum.\\
   From visual inspection, we decided to exclude from the subsequent chemical analysis: four stars (179$\_$NDC, 201$\_$NDC, 37024$\_$J10 and 48099$\_$J10) with $T_{\mathrm{eff}}<3900$ K because their spectra were contaminated by the TiO molecular bands, and one star (371$\_$NDC) with a very low SNR with respect to other stars with similar magnitude.

\section{CHEMICAL ANALYSIS} \label{sec:chem}
   \subsection{Atmospheric parameters}
\begin{figure}
   \centering 
   \includegraphics[width=\columnwidth]{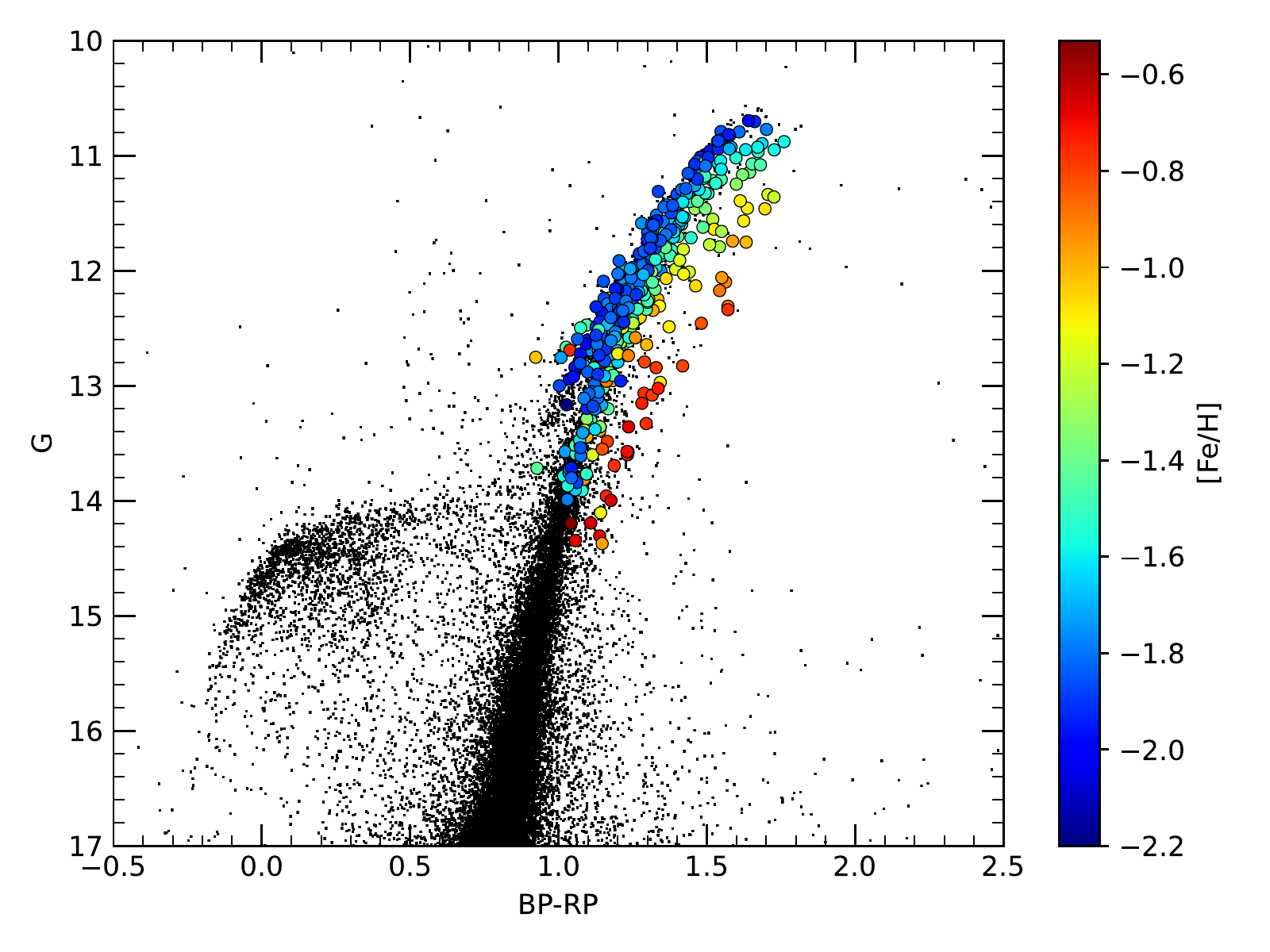}
    \caption{Color-magnitude diagram of $\omega$ Centauri. Black points represent all the targets of $\omega$ Centauri observed with GAIA, while the targets selected for this study are colored according with their metallicity. The color scale is shown on the right side.}
    \label{cmd}
   \end{figure}

   We derived the stellar parameters from Gaia early Data Release 3 photometry (\citealt{Gaia_16}, \citealt{Gaia_21}). Figure $\ref{cmd}$ shows the color-magnitude diagram of $\omega$ Centauri where the position of the spectroscopic targets is marked.\\
   Effective temperatures ($T_{\mathrm{eff}}$) were computed using the empirical $(BP-RP)_0-T_{\mathrm{eff}}$ relation by \citet{mucc_21}, based on the InfraRed Flux Method. The dereddened color $(BP-RP)_0$ was obtained by assuming a color excess of $E(B-V)=0.12\pm0.02$ (\citealt{harris10}) and adopting an iterative procedure following the scheme proposed by \citet{Gaia_2018}. Internal errors in $T_{\mathrm{eff}}$ due to the uncertainties in photometric data, reddening and $(BP-RP)_0-T_{\mathrm{eff}}$ relation are of the order of 85-115 K.\\
   Surface gravities ($\log g$) were obtained from the Stefan-Boltzmann relation using the photometric $T_{\mathrm{eff}}$ and assuming a typical mass of $0.80\;\mathrm{M_{\odot}}$. Luminosities were computed using the dereddened G-band magnitude with the bolometric corrections from \citet{andrae_18} and a true distance modulus DM$_0=13.70\pm0.06$ (\citealt{principe06}). We computed the uncertainties in gravities by propagating the uncertainties in $T_{\mathrm{eff}}$, distance modulus and photometry. These uncertainties are of the order of $0.1$ dex.\\
   Microturbolent velocities ($v_t$) were obtained adopting the relation between $v_t$ and $\log\;g$ by \citet{kirby09}. This relation provides values of $v_t$ of about 1.6-2.0 km s$^{-1}$. To compute the uncertainties in $v_t$ we assumed a conservative error of $0.2$ km s$^{-1}$.  \\
   The derived atmospheric parameters for all the analysed targets are listed in Table $\ref{tab:1}$, together with some additional information. 
   
   \begin{deluxetable*}{cccccccccc}
   \tablenum{1}
   \tablecaption{Data for the analysed targets in $\omega$ Centauri. \label{tab:1}}
   \tablewidth{0pt}
   \tablehead{
   ID & G & $v_r$ & $T_{\mathrm{eff}}$ & $\log g$ & $v_t$ & [Fe/H] & [Na/Fe] & [Mg/Fe] & [K/Fe]\\
       & (mag) & (km s$^{-1}$) & (K) & (dex) & (km s$^{-1}$) & - & - & - & - 
   }
   \startdata
      48$\_$NDC & 10.7035 & 220.6 & 4041 & 0.44 & 2.03 & $-1.92\pm 0.07$ & $-0.01\pm0.06$ & $0.42\pm 0.04$ & $0.56\pm 0.15$\\
      74$\_$NDC & 11.0120 & 215.7 & 4273 & 0.72 & 1.96 & $-1.93\pm 0.10$ & $0.21\pm0.04$ &  $0.44\pm 0.03$ & $0.41\pm 0.11$ \\
      84$\_$NDC & 10.9489 & 220.3 & 3971 & 0.49 & 2.02 & $-1.57\pm 0.07$ & $0.08\pm0.09$ & - & $0.31\pm 0.17$ \\
      161$\_$NDC & 11.2326 & 247.3 & 4301 & 0.82 & 1.94 & $-1.74\pm 0.10$ & $-0.26\pm0.07$ &  $0.42\pm 0.02$ & $0.29\pm 0.11$ \\
      182$\_$NDC & 11.3292 & 207.3 & 4247 & 0.83 & 1.94 & $-1.53\pm 0.07$ & $-0.06\pm0.05$ &  $0.51\pm 0.05$ & $0.25\pm 0.14$ \\
      357$\_$NDC & 11.7420 & 230.9 & 4153 & 0.93 & 1.92 & $-0.97\pm 0.06$ & - & - & $0.39\pm 0.18$  \\
      480$\_$NDC & 12.2484 & 226.7 & 4503 & 1.35 & 1.82 & $-1.02\pm 0.11$ & - & - & $0.48\pm 0.13$ \\
      27048$\_$J10 & 11.8638 & 242.0 & 4529 & 1.21 & 1.85 & $-1.59\pm 0.10$ & $0.37\pm0.06$ & $-0.07\pm 0.05$ & $0.53\pm 0.11$ \\
      27094$\_$J10 & 12.8344 & 237.9 & 4778 & 1.73 & 1.73 & $-1.83\pm 0.11$ & $0.24\pm0.06$ & $0.44\pm 0.06$ & $0.14\pm 0.08$ \\
      29085$\_$J10 & 12.3903 & 226.7 & 4730 & 1.53 & 1.78 & $-1.75\pm 0.10$ & $0.21\pm0.06$ & $-0.37\pm 0.07$ & $0.52\pm 0.10$ \\
   \enddata
   \tablecomments{This is a portion of the entire table.}
   \end{deluxetable*}

   \subsection{Line lists and tools}
   The chemical analysis was performed using one-dimensional, Local Thermodynamic Equilibrium (LTE), plane-parallel geometry model atmospheres computed with the code ATLAS9 (\citealt{castelli04}) that treats the line opacity through the opacity distribution functions (ODF) method. All the models are calculated using the ODFs computed by \citet{castelli04} with $\alpha$-enhanced chemical composition and without the inclusion of the approximate overshooting in the calculation of the convective flux. \\
   A line list of relatively strong and unblended spectral features at the resolution of our GIRAFFE observations was selected from the Kurucz/Castelli line lists\footnote{\url{https://wwwuser.oats.inaf.it/castelli/linelists.html}} by comparing observed spectra with synthetic ones having appropriate metallicity and $T_{\mathrm{eff}}$. Model spectra were calculated with the SYNTHE code in its Linux version (\citealt{sbordone04}, \citealt{kurucz05}). \\
   We derived the chemical abundances of Fe, Na, Mg and K from the comparison between measured and theoretical Equivalent Widths (EWs) with the package GALA (\citealt{gala}). EWs were measured with the code DAOSPEC (\citealt{stetson08}), through the wrapper 4DAO (\citealt{4dao}). \\
   Non-LTE (NLTE) corrections for Na (at 5682 and 5688 $\angstrom$) and K (at 7699 $\angstrom$) were calculated by interpolating into the grids of \citet{lind_11} and \citet{reggiani_19}, respectively. \\
   Solar abundances are from \citet{grev_98}.\\
   
   \subsection{Error estimates}
   Uncertainties associated with the chemical abundances were calculated as the sum in quadrature of the error related to measurement process and the errors associated to the atmospheric parameters\footnote{If we add in quadrature all the possible sources of errors the uncertainties would increase from a minimum of 0.02 for Na to a maximum of 0.06 dex for K, which is the most problematic element to measure. However, the results of our study are the same even if we adopt this extremely conservative approach to estimate errors.}.\\
   Error related to the measurement was calculated as the line-to-line scatter divided by the square root of the number of used lines. When only one line was present the error was calculated by varying the EW of $1\sigma_{EW}$ (i.e. the EW error provided by DAOSPEC).\\
   Errors related to the adopted atmospheric parameters were calculated by varying only one parameter at time, keeping the others fixed to their best value, and recalculating each time the chemical abundances. At the end all the error sources are added in quadrature. This approach is the most conservative in the calculation of uncertainties because it does not take into account the correlation terms between parameters. So, it should be regarded as an upper limit to the real error associated to the measurements.\\
   Since the abundances are expressed as abundance ratios, the total uncertainties in [Fe/H] and [X/Fe] are calculated as follows:
   \begin{equation}
       \sigma_{[Fe/H]}=\sqrt{\frac{\sigma_{Fe}^2}{N_{Fe}}+(\delta_{Fe}^{T_{\mathrm{eff}}})^2+ (\delta_{Fe}^{\log g})^2 + (\delta_{Fe}^{v_t})^2}
   \end{equation}
   
   \begin{myequation}
   \begin{split}
       & \sigma_{[X/Fe]}   =  \\
       & = \sqrt{\frac{\sigma_{X}^2}{N_{X}} + \frac{\sigma_{Fe}^2}{N_{Fe}} + (\delta_{X}^{T_{\mathrm{eff}}} - \delta_{Fe}^{T_{\mathrm{eff}}})^2 + (\delta_{X}^{\log g} - \delta_{Fe}^{\log g})^2 + (\delta_{X}^{v_t} - \delta_{Fe}^{v_t})^2}
   \end{split}
   \end{myequation}
   where $\sigma_{X,Fe}$ is the line-to-line scatter, $N_{X,Fe}$ the number of lines used to compute the abundance and $\delta_{X,Fe}^i$ are the abundance variations obtained after variation of the atmospheric parameter $i$. 

\section{RESULTS} \label{sec:results}

   \subsection{Mg-K anti-correlation}
\begin{figure*}
   \centering
   \includegraphics[width=\textwidth]{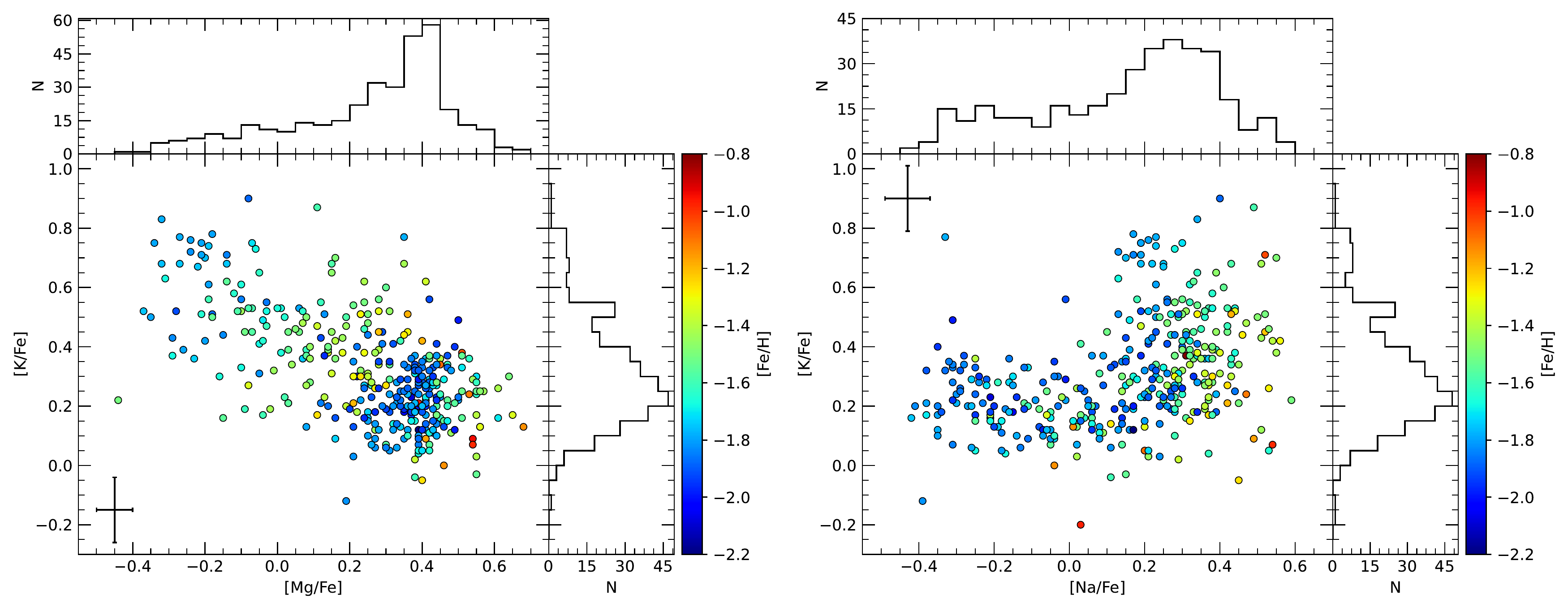}
      \caption{Left panel: behaviour of [K/Fe] as a function of [Mg/Fe]. Each star is color coded according to its value of [Fe/H] and the color scale is shown on the right side. The error bar represents the typical error associated to the abundance ratios. The distribution of [K/Fe] and [Mg/Fe] are shown as marginalized histograms.\\
      Right panel: the same as the left panel, but for [K/Fe] and [Na/Fe] abundance ratios.}
         \label{2_panels_hist}
   \end{figure*}

   Potassium elemental abundances were derived for a total of 440 stars. Moreover, we obtained Fe, Na and Mg abundances for 440, 359 and 357 stars, respectively (Table $\ref{tab:1}$). \\
   The mean value of the [K/Fe] distribution is $+0.31$ dex ($\sigma=0.19$ dex), with values ranging from $-0.20$ dex up to $+0.94$ dex. In this calculation we excluded one star showing a significant higher [K/Fe] and that we will discuss in Section $\ref{K_rich_sec}$.\\
   The main result of our study is the presence of a clear anti-correlation between [Mg/Fe] and [K/Fe] shown in the left-hand panel of Figure $\ref{2_panels_hist}$. We identified a sample of stars with [Mg/Fe]$<0.0$ dex that are systematically enriched in [K/Fe]. This kind of stars are rarely observed in GCs.\\
   Also K correlates with Na (see the right-hand panel of Figure $\ref{2_panels_hist}$). 
   To quantitatively assess these results, we computed for the couples [K/Fe]-[X/Fe] (with X = Na and Mg) the Spearman's correlation coefficient ($C_S$) and the corresponding two-tailed probability that an absolute value $C_S$ larger than the observed one can derive from non-correlated random variables. We found $C_S= +0.41$ and $-0.52$ for [K/Fe]-[X/Fe] (with X = Na and Mg) respectively, leading to null probability that the observed correlations arose by chance from uncorrelated variables.  \\
   A hint of the presence of a Mg-K anti-correlation in $\omega$ Centauri was firstly detected by \citet{meszaros20}. They found seven stars with [Mg/Fe]$<0.0$ dex (out of a total sample of 898 stars) in which [K/Fe] seems to be enhanced with respect to stars with [Mg/Fe]$>0.0$ dex (see their Figure 10). Unfortunately, the K lines in H band measured by \citet{meszaros20} are weak and blended with other lines. Therefore, the authors concluded that it is not possible to claim convincingly the presence of a K enhancement in the Mg-poor stars of $\omega$ Centauri. Hence, this is the first time that the presence of a strong Mg-K anti-correlation is undoubtedly established in $\omega$ Centauri.
   
   \subsection{A super K-rich star}\label{K_rich_sec}
\begin{figure*}
   \centering
   \includegraphics[width=\textwidth]{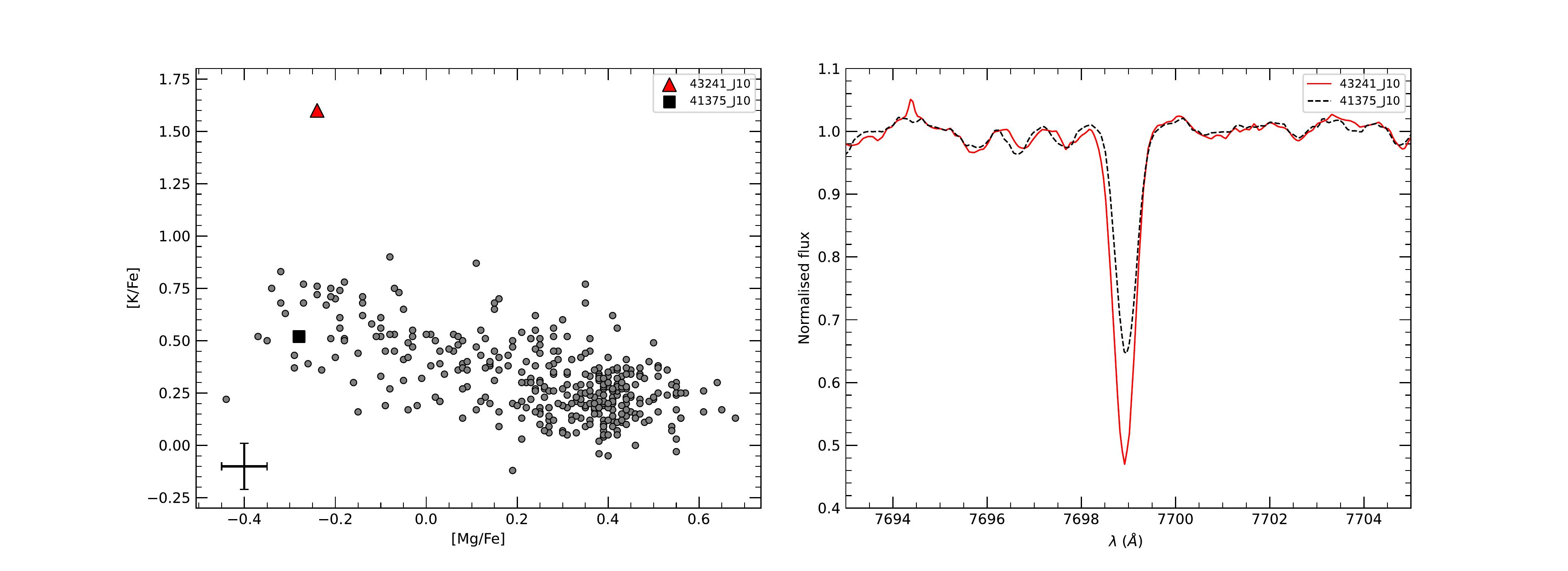}
      \caption{Left panel: behaviour of [K/Fe] as a function of [Mg/Fe] with the inclusion of the K-rich star 43241$\_$J10 (red triangle). The black square represents the reference star 41375$\_$J10 with similar atmospheric parameters as the K-rich star.\\ 
      Right panel: comparison between the GIRAFFE spectra of the two stars around the strong K line at 7699 $\angstrom$. The large K enhancement of 43241$\_$J10 is clearly visible from this comparison.}
         \label{2panels}
   \end{figure*}

   In the analysed sample we identified a peculiar star, named 43241$\_$J10, that clearly stands out from the mean locus of other $\omega$ Centauri members in the [Mg/Fe] vs [K/Fe] distribution, as can be seen in the left-hand panel of Figure $\ref{2panels}$. Indeed, this star has [K/Fe]$=+1.60$ dex. \\
   We compared its spectrum with that one of a reference star (41375$\_$J10), with similar atmospheric parameters and [Fe/H] and [Mg/Fe] abundance ratios. In the right-hand panel of Figure $\ref{2panels}$ the observed K lines of the two stars are directly compared, showing the obvious difference in line depth, implying intrinsic difference in the [K/Fe] abundance. \\
   This star was also included in the sample studied by \citet{johnson}. For the stars in common between the two studies we found that our temperatures are on average ($140\pm94$ K) higher than those computed by \citet{johnson}. Thus, the difference of $+185$ K computed for 43241$\_$J10 in this study with respect to \citet{johnson} is well within the mean difference between the two temperature scales. Even if we adopt the temperatures by \citet{johnson}, the difference in K abundance of 43241$\_$J10 with respect to the other stars remains. \\
   Therefore, we can conclude that the high [K/Fe] abundance of this star is real and not an artifact of the analysis.

\section{DISCUSSION} \label{sec:disc}
\begin{figure}
   \centering
   \includegraphics[width=\columnwidth]{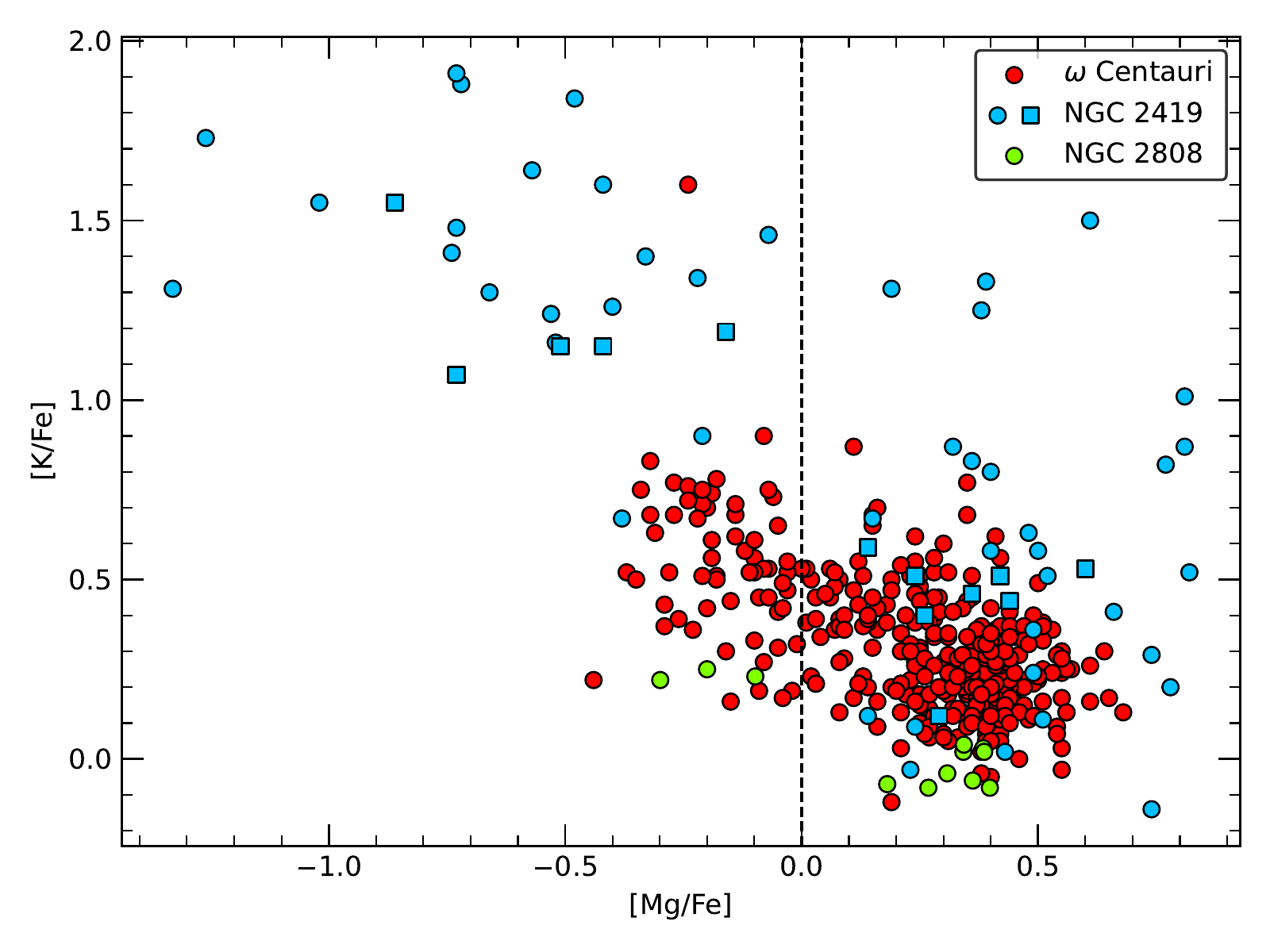}
      \caption{Run of the [K/Fe] abundance against [Mg/Fe] for stars in GCs. $\omega$ Centauri stars are shown as red circles. Stars in NGC 2808 are plotted as green circles (\citealt{mucc15}), whereas stars in NGC 2419 are in blue (data from \citet{mucc12} and \citet{cohen12} are shown as circles and squares respectively). The dashed line splits the Mg-poor from the Mg-rich stars.}
         \label{Mg_K_clusters}
   \end{figure}

   Our analysis of a large sample of $450$ giants in $\omega$ Centauri has revealed the presence of (1) a large intrinsic spread in the [K/Fe]; (2) a correlation between the K abundances and other light elements (i.e. Na and Mg) which are observed to vary in GCs showing MPs. In particular, we detected the presence of a prominent Mg-K anti-correlation. This finding makes $\omega$ Centauri the third stellar system after NGC 2419 (\citealt{mucc12} and \citealt{cohen12}) and NGC 2808 (\citealt{mucc15}) in which a sub-population of stars with [Mg/Fe]$<0.0$ dex and enriched in K is present (Figure $\ref{Mg_K_clusters}$). \\
   The amplitude of the [K/Fe] spread in $\omega$ Centauri is intermediate between those of NGC 2419 and NGC 2808. \citet{carretta21} suggested the presence of a weak (but statistically significant) Mg-K anti-correlation in NGC 4833. In this cluster the stars showing the highest Mg depletion exhibit a typical overabundance in K of 0.1 dex with respect to the most Mg-rich stars. However, this cluster does not harbor stars with [Mg/Fe]$<0.0$ dex at variance to NGC 2419, NGC 2808 and $\omega$ Centauri. \\
   The chemical complexity of the stellar populations of $\omega$ Centauri deserves further discussion of the measured [K/Fe] abundance ratios and their correlations with other light elements for the different groups of stars with distinct metallicities. The extent of the Mg-K anti-correlation is mainly driven by the metal-poor ([Fe/H]$\leqslant-1.70$ dex) component which exhibits a bimodal distribution both in [K/Fe] and in [Mg/Fe]. In particular, the group of metal-poor stars with [Mg/Fe]$<0.0$ dex has a mean value of [K/Fe] $\sim 0.4$ dex higher than that of the Mg-rich, metal-poor stars. On the other hand, an evident but less extended Mg-K anti-correlation is detected also among the stars with intermediate [Fe/H] ($-1.70<$[Fe/H]$\leqslant-1.30$ dex). Finally, for the most metal-rich population ([Fe/H]$>-1.30$ dex) we cannot conclude for the presence of a Mg-K anti-correlation. For this group of stars we do not have objects with [Mg/Fe]$<0.0$ dex and the spread in [K/Fe] is less extended than that of the other sub-populations. \\
   The evidence of a correlation between K and the other light elements involved in MP phenomenon supports the idea that the spread in [K/Fe] can be ascribed to the self-enrichment process typical of GCs. \\
   The detection of an intrinsic variation in K abundances in the most massive clusters represents a serious challenge for theoretical models for the MPs. In the model proposed by \citet{ventura12} to explain the Mg-K anti-correlation in NGC 2419 and based on AGB and super-AGB stars, the production of K occurs during the HBB phase, by proton capture reaction on Argon nuclei at temperatures $T\sim 10^8$ K. Even if AGB yields discussed by \citet{ventura12} are qualitatively able to explain a Mg-K anti-correlation, the observed [K/Fe] abundance ratios in the extreme case of NGC 2419 can be reproduced only with an increase (1) in the cross section of the reaction $^{38}\textrm{Ar}(p,\gamma)^{39}\textrm{K}$ by a factor 100 with respect to literature or (2) in the temperature at the base of envelope up to $1.5\cdot 10^8$ K during the HBB. Moreover, \citet{ventura12} also predicted that Mg-poor stars would show normal Na abundances, if the Mg-K anti-correlation was indeed produced by AGB and super-AGB stars. This is not observed in $\omega$ Centauri, where the Mg-poor stars are also enhanced in Na.\\
   On the other hand, \citet{pran17} discuss the possible production of K from FRMS and supermassive stars, ruling out both the classes of polluters because not able to reach the temperature of K-burning. Therefore, the existence of MPs in GCs remains largely unexplained.\\
   Another interesting result that we found is the presence of a peculiar star with a [K/Fe]$\sim 1$ dex higher than the abundances of Mg-poor, metal-poor stars. Even if the origin of this extraordinary overabundance of [K/Fe] is unclear, it is worth noting that the Mg and K abundances of this star are very similar to those of NGC 2419 (see Figure $\ref{Mg_K_clusters}$) for which \citet{ventura12} proposed that the K enhanced sub-population was born from the AGB and super-AGB ejecta without dilution process.\\
   The metal-poor population of $\omega$ Centauri exhibits the usual light element anti-correlations typical of GCs with MPs. Assuming a scenario where the K enhancement in Mg-poor stars is due to pollution by the first generation of AGB and super AGB stars, the ejected gas undergoes a dilution process with the pristine GC gas before forming new generations of stars (\citealt{dercole_08}, \citeyear{dercole_10}). In this scenario is possible that a small fraction of stars will form directly from the ejecta of polluters without dilution with pristine material. In this framework, we can suppose that the super K-rich star we discovered formed directly from the pure ejecta of AGB stars before dilution, while the other Mg-poor stars show a lower [K/Fe] due to a level of dilution with pristine material. Further inspections of this star are necessary to understand its origin. In particular, it would be useful to study other light elements involved in the proton-capture reactions (C, N, O, etc).\\
   In conclusion, with the present analysis we support the idea that the observed spread in [K/Fe] in $\omega$ Centauri stars is associated to a self-enrichment process typical of GCs. So far a clear Mg-K anti-correlation was found only in three GCs, namely $\omega$ Centauri, NGC 2808 and NGC 2419. Those clusters are among the most massive stellar systems of the Milky Way, so we suggest that this anomaly is a manifestation of an extreme self-enrichment process that occurs only in the most massive and/or metal-poor clusters.

\begin{acknowledgements}
We are grateful to the anonymous referee for his/her useful suggestions. \\
CL acknowledges funding from Ministero dell’Università e della Ricerca (MIUR) through the Programme ‘Rita Levi Montalcini’ (grant PGR18YRML1).
\end{acknowledgements}

\bibliography{biblio}{}
\bibliographystyle{aasjournal}

\end{document}